# Adsorption and Diffusion of $F_2$ molecules on Pristine Graphene

Yong Yang[2, 1*], Fuchi Liu[1] and Yoshiyuki Kawazoe[3,4]

[1]*College of Physics and Technology, Guangxi Normal University, Guilin 541004, China.*

[2]*Key Laboratory of Materials Physics, Institute of Solid State Physics, Chinese Academy of Sciences, Hefei 230031, China.*

[3]*New Industry Creation Hatchery Center (NICHe), Tohoku University, 6-6-4 Aoba, Aramaki, Aoba-ku, Sendai, Miyagi 980-8579, Japan.*

[4]*Department of Physics and Nanotechnology, SRM Institute of Science and Technology, Kattankulathurm, 603203, TN, India.*

**Abstract:**

The adsorption and diffusion of $F_2$ molecules on pristine graphene have been studied using first-principles calculations. For the diffusion of $F_2$ from molecular state in gas phase to the dissociative adsorption state on graphene surface, a kinetic barrier is identified, which explains the inertness of graphene in molecular $F_2$ at room temperature, and its reactivity with $F_2$ at higher temperatures. Studies on the diffusion of $F_2$ molecules on graphene surface determine the energy barriers along the optimal diffusion pathways, which help to understand the high stability of fluorographene.

*Corresponding Author. E-mail: wateratnanoscale@hotmail.com; yyang@theory.issp.ac.cn

# 1. Introduction

Chemical modification is an effective method of tailoring the physical and chemical properties of graphene [1-5]. With the decoration of foreign species of atoms or molecules, functional derivatives of graphene can be created. To date, at least three typical derivatives of graphene are reported, namely: Graphene Oxide (GO), which is obtained by the decoration of hydroxyl and epoxy groups on graphene; Graphane, an extended two-dimensional hydrocarbon with one-to-one molar ratio of carbon and hydrogen atoms, which was predicted firstly by first-principles calculations [6] and synthesized experimentally later [2]; and Fluorographene, which was synthesized recently, is a two-dimensional counterpart of Teflon with one-to-one molar ratio of carbon and fluorine atoms [3]. It is found by experiment that hydrogenated graphene can rapidly lose the adsorbed H atoms at moderate temperatures [2], which casts doubt on the realistic application of graphane where the thermal stability is a prerequisite. On the contrary, due to the much stronger F-C bond (with comparison to H-C bond of graphane), fluorographene is observed to be inert and stable in air up to ~ 400 °C [3].

Fluorination of graphene can open a finite gap in the energy band structure and therefore tune its electronic and optical properties from the original metallic state to semiconducting and even insulating state [3, 7]. Furthermore, it was shown by experimental [8] and theoretical studies [9] that local magnetic moment may appear in fluorinated graphene, resulting in the so-called $d^0$ magnetism [10]. It is shown recently that fluorination can tune the electronic and optical properties of the other two-dimensional (2D) systems such as bilayer graphene [11] and 2D-SiC [12]. On the other hand, pristine graphene is found to be stable at room temperature in the presence of the $F_2$ molecules [13]. Such inertness is surprising when considering the extremely strong oxidizing characteristics of $F_2$. However, for decades, the underlying mechanism remains unclear. In this work, we attempt to resolve this puzzle from the atomic level using first-principles calculations. We identify a kinetic barrier for the diffusion and adsorption of a $F_2$ molecule from the molecular state in gas phase to the atomic adsorption state on graphene surface. We have further studied the energy pathway for the diffusion of the adsorbed $F_2$ molecule on graphene, along which the key energy barriers are determined. The existence of such energy barriers of diffusion helps to understand the following experimental observations: 1) At the presence of molecular $F_2$, the inertness of graphene at room temperature and its reactivity at moderately higher temperatures; and 2) The high thermal stability of fluorographene.

## 2. Theoretical Methods

All the calculations were carried out by using the Vienna *ab initio* simulation package (VASP) [14, 15], which is based on density functional theory (DFT). The electron wave function and the electron-ion interactions were respectively described by a plane wave basis set and the projector-augmented-wave (PAW) potentials [16, 17]. The exchange-correlation interactions of electrons were described by the PBE type functional [18]. The energy cutoff for plane waves is 600 eV. For the structural relaxation and total energy calculation of the $F_2$/graphene system, an $8\times8\times1$ Monkhorst-Pack k-mesh [19] was generated for sampling the Brillouin zone (BZ). The graphene sheet on which a $F_2$ molecule is adsorbed and diffuses is modeled by a ($5\times5$) supercell of graphene which extends periodically in the *x*- and *y*-direction, separated by a vacuum layer of ~ 15 Å in the *z*-direction to minimize the artificial interactions due to periodic boundary condition employed in the simulations. The adsorption energy of a $F_2$ molecule is calculated as follows:

$E_{ads} = E[\text{graphene}] + E[(F_2)_{\text{isolated}}] - E[F_2/\text{graphene}]$ (1), where the three terms $E[F_2/\text{graphene}]$, $E[\text{graphene}]$, and $E[(F_2)_{\text{isolated}}]$ are respectively the total energies of the adsorption system, the graphene substrate, and an isolated $F_2$ molecule. To study the diffusion of a $F_2$ molecule from the gas phase to its adsorption state on graphene, and to study its diffusion process on the graphene surface, we employed the nudged elastic band (NEB) method [20, 21] implemented in VASP [14, 15] to locate the saddle points of potential energy surface and search for the minimum energy pathway of diffusion. The number of images (intermediate states) considered in the NEB method is eight in studying the dissociation process of the $F_2$ molecule, and is four in studying the diffusion of the $F_2$ molecule/atoms from one stable adsorption site

## 3. Results and Discussion

Shown in Fig. 1, are a number of typical adsorption configurations of one $F_2$ molecule (molecular and dissociative states) on graphene. The calculated adsorption energy is ~ 1.81 eV, 1.65 eV, 1.03 eV and 1.16 eV for Configurations I, II, III, and IV, respectively. In contrast to the chemisorption of Configurations I to IV, the $F_2$ molecule in Configuration V is physically absorbed on the graphene sheet, with the adsorption energy of ~ 0.40 eV. The adsorption energies and the related parameters describing the adsorption geometries are listed in Table 1. It is clear that Configuration I, where the two F atoms of the $F_2$ molecule are separately adsorbed on the top sites of two C atoms which are the $3^{rd}$ nearest neighbors of each other (Figs. 1(a), 1(c)), is energetically the most stable. The second stable is Configuration II, in which the two F atoms stay atop two C atoms which are the $1^{st}$ nearest neighbors of each other. In gas phase, our calculations predict a F-F bond length of ~ 1.42 Å and bond energy of ~ 2.71 eV, which are comparable with previous DFT-PBE calculations (1.41 Å, 2.35 eV) [22].

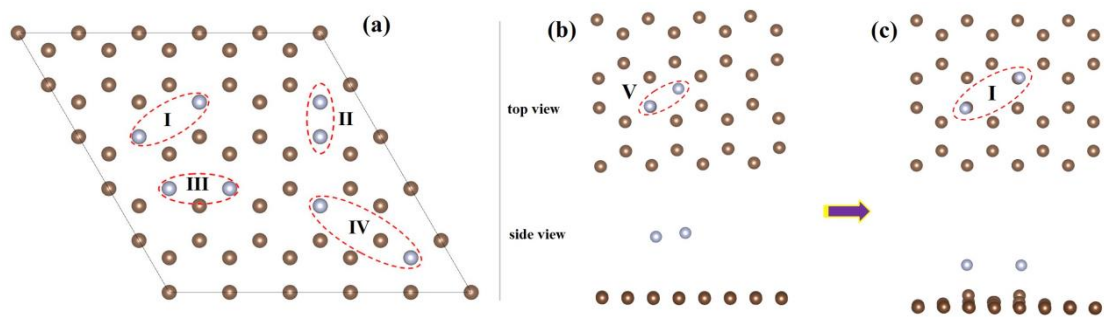

**Fig. 1.** (a) Schematics of some top-site adsorption configurations (I to IV) of a $F_2$ molecule on graphene. (b) and (c): Top and side views of adsorption Configuration V (b) and Configuration I (c), where the transition is indicated by an arrow. For the clarity of illustration, only part of the (5×5) supercell is shown in panels (b) and (c).

Compared with the molecules in gas phase, a physically adsorbed $F_2$ molecule stays above the graphene sheet at a height of ~ 3 Å (Fig. 1(b), Configuration V), with a slightly elongated bond length (~ 1.64 Å, see Table 1). Considering the adsorption energy and the slight change in molecular structure, it is evident that a $F_2$ molecule will attach spontaneously from the gas phase to graphene. Therefore, Configuration V serves as a good starting point for studying how a $F_2$ molecule diffuses and dissociates on the graphene surface, and finally reaches the most stable Configuration I, as schematically depicted in Figs. 1(b)-(c).

Figure 2 shows the optimal energy pathway for the diffusion and dissociation of a $F_2$ molecule from the initial molecular state (Configuration V) to the final dissociated state on graphene (Configuration I). Using the NEB method [18, 19], we are able to locate some typical transition states along the diffusion pathway, as marked by capital letters A-D in Fig. 2. The dissociation of the $F_2$ molecule is evidenced by the elongation of the bond length and the adsorption is indicated by the decrease of its distance to the graphene surface, d($F_2$-Gr), as shown in Fig. 2.

An energy barrier on the diffusion pathway from Configuration V to the transition state B, $E_b$ ~ 0.33 eV, is identified. It would be instructive to have a discussion on the reaction rate of the transition from Configuration V to Configuration I. A common method for estimation of the reaction rate is the Arrhenius equation: $K = K_0 e^{-\Delta E/(k_B T)}$, where $K$ is the rate constant, $K_0$ is the prefactor, $\Delta E$ is the activation energy, $k_B$ is the Boltzmann constant, and $T$ is the temperature. This empirical formula works well for reactions in homogeneous systems such as gas phase. However, severe problem may be encountered in inhomogeneous systems such as dynamical processes (including reactions) on surfaces. Previous works [23-25] have demonstrated that the experimentally fitted value of $K_0$ can be much smaller or larger than the "normal value", which is ~ $10^{12}$ s$^{-1}$. For instance, in the process of hydrogen desorption from amorphous hydrogenated silicon, the deduced prefactor $K_0$ from experimental data can differ by 14 orders of magnitude (10 s$^{-1}$ ~ $10^{15}$ s$^{-1}$) [23]. On the other hand, the value of $K_0$ is affected by many factors (available surface sites, recombination, gas pressure, substrate temperature, translation and rotation motions of the adsorbates, …), which make it difficult for theoretical evaluation of $K_0$ with reference to experimental data. Therefore, our discussion will focus on the equilibrium probability of the transition states, $e^{-\Delta E/(k_B T)}$, which is well established by experiments.

The diffusion from transition state B to the final state (Configuration I) is expected to be spontaneous due to the downhill characteristics in the energy landscape. At room temperature ($T \sim 300$ K), the probability of surmounting such a kinetic barrier ($\Delta E = E_b$) is therefore estimated as $p \sim e^{-E_b/(k_B T)} \sim 2.86 \times 10^{-6}$. At $T \sim 350$ K ($\sim 77$ °C), the probability is $p \sim 1.77 \times 10^{-5}$, increased by about one order of magnitude, at which detectable reaction between $F_2$ molecules and graphene comes into play, though a longtime of exposure is required to complete the reaction [3]. When the temperature is increased to $\sim 480$ K ($\sim 207$ °C), the probability is $p \sim 3.43 \times 10^{-4}$, about two order of magnitude as larger than that at room temperature. In fact, the significant increase in the reaction rate as expected by theoretical calculation has been experimentally demonstrated by the fluorination process of graphene, in which the reaction time was reduced from weeks to hours when the sample was prepared at $\sim 200$ °C [3]. The existence of a kinetic barrier of diffusion explains why graphene is inert and stable in the atmosphere of $F_2$ at room temperature, and why it readily reacts with $F_2$ and gets fluorinated at temperatures well above 300 K.

Based on the time of fluorination reaction provided by the experimental data [3], we can estimate the reaction rate $r$, rate constant $K$, and consequently the prefactor $K_0$ of the Arrhenius equation. To the first order, the reaction rate is $r = K\theta$, where $\theta$ is the probability of finding the 3$^{rd}$ nearest neighboring (3-NN) C sites for $F_2$ adsorption. Approximately, the value of $\theta$ can be calculated as the ratio of the solid angle occupied by the dissociated $F_2$ molecule (Configuration I) to the whole space: $\theta = \frac{1}{4\pi} \times \Omega \times 3$. The factor 3 corresponds to the three equivalent 3-NN sites around one C atom. In our case, $\Omega = \frac{\pi r_F^2}{R_{C-C}^2}$, where $r_F$ is the radius of F atom, which is $\sim 1.1$ Å, and $R_{C-C} \sim 2.839$ Å, is the distance between every two C atoms at the 3-NN sites. It follows that $\theta = \frac{3}{4} \times (\frac{r_F}{R_{C-C}})^2 \sim 0.11$. Assuming that $K_0 \sim 1$s$^{-1}$, then $r = e^{-E_b/(k_B T)} \times 0.11$. At $T \sim 480$ K, $r \sim 3.773 \times 10^{-5}$s$^{-1}$, the reaction time: $t_r \sim 1/r \sim 26504$ s $\sim 7.36$ hours. The reaction time at $T \sim 350$ K, 300 K is similarly estimated to be $t_r \sim$ 142.67 hours (~ 5.94 days), ~ 882.96 hours (~ 36.79 days, long enough to be considered as stable), respectively. The estimated time scales are in very good agreement with the orders of magnitude reported in experiment [3]. Therefore, the prefactor $K_0$ in the fluorination reaction is estimated to be $K_0 \sim 1s^{-1}$. The "abnormally" small value of the prefactor $K_0$ may be explained using the theory of large energy fluctuation of small number of atoms in condensed phase, which was suggested to understand the abnormal kinetic process in Si-based systems [23]. In our case, the large energy fluctuation of neighboring atoms at transient time scale ($\sim 10^{-13} - 10^{-12}$ s) would distort the local atomic structures and elevate the transient kinetic barrier for the dissociation of the adsorbed $F_2$. For instance, at $T = 300$ K, if the transient kinetic barrier is enlarged by ~ 0.3 eV ($\Delta E_b \sim 0.3$ eV, and $E_b \sim 0.6$ eV), then the corresponding $K_0$ will be the normal value (~ $10^{12} s^{-1}$, i.e., atomic vibrational frequency) for the same rate constant. The estimated small value of $K_0$ in this work is actually the time-average of the kinetic process at equilibrium state, where the transient atomic vibration and energy fluctuation cancels with each other. As a result, the transient larger values of $K_0$ and $E_b$ are averaged out to be much smaller values which are estimated at equilibrium state and macroscopic time scale.

**Table 1.** The adsorption energies and geometric parameters [the F-C distances: d_F-C (two values), and the F-F bond lengths: d_F-F] describing the top-site adsorption configurations of a $F_2$ molecule on graphene.

| Configuration | $E_{ads}$ (eV) | d_F-C (Å) | d_F-F (Å) |
|---|---|---|---|
| I | 1.81 | 1.50; 1.50 | 3.06 |
| II | 1.65 | 1.46; 1.46 | 2.38 |
| III | 1.03 | 1.55; 1.58 | 4.45 |
| IV | 1.16 | 1.56; 1.56 | 2.94 |
| V | 0.40 | 3.00 | 1.64 |
| $F_2$ in gas phase | --- | --- | 1.42 |

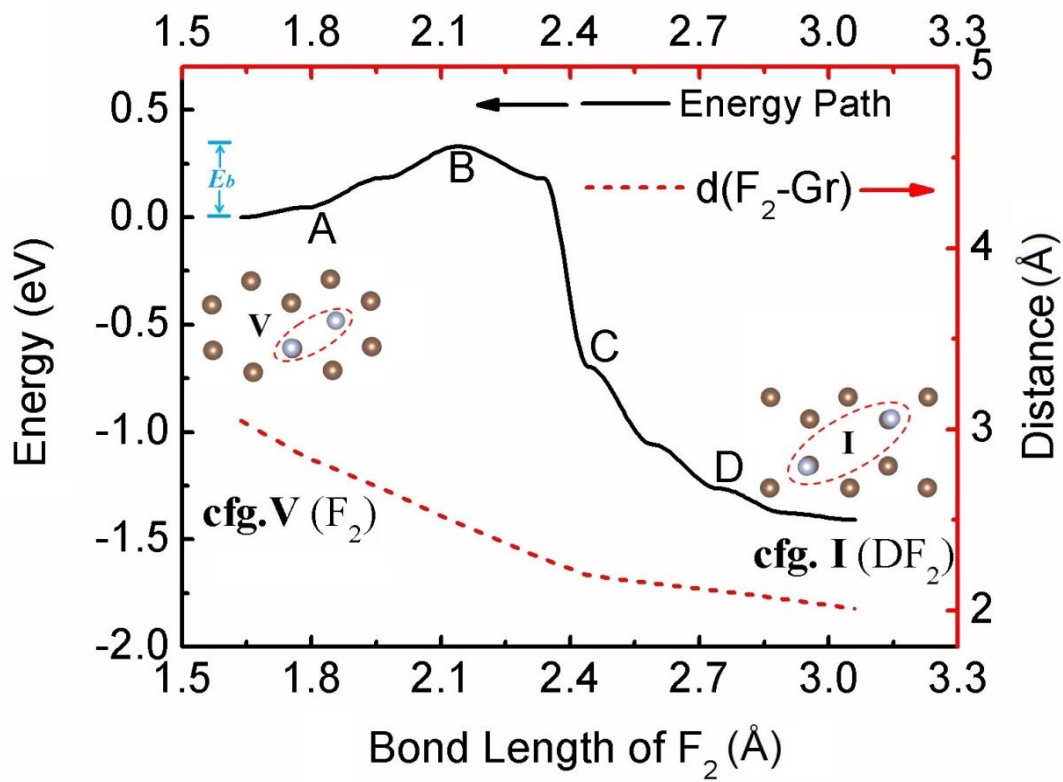

**Fig. 2.** The calculated energy pathway (solid line) along which a $F_2$ molecule diffuses from molecular state to dissociative state ($DF_2$); and the distance (dashed line) from the geometric center of the $F_2$ molecule to the graphene surface, denoted by $d(F_2\text{-Gr})$. "Configuration" is abbreviated as cfg, and "graphene" is abbreviated as Gr.

As seen from Fig. 2, the bond length of the $F_2$ molecule gradually increases from the initial value of ~ 1.64 Å to the final value of ~ 3.06 Å. During the process of bond length elongation, the distance of the $F_2$ molecule to the graphene sheet decreases quickly and continuously, from the initial value of ~ 3 Å to ~ 2.2 Å at the transition state C, and decreases slightly to ~ 2 Å when arriving at the final state. The atomic configurations for the typical transition states (A, B, C, and D in Fig. 2) are schematically shown in Fig. 3, where the elongation of the $F_2$ bond length and the decrease of $F_2$ – graphene distance are clearly illustrated. We have further studied the electron transfer between the $F_2$ molecule and the graphene substrate, by doing Bader analysis [26, 27] on the charge density of the adsorption systems: the initial molecular state (cfg. V), the transition states (A, B, C, D), and the final dissociative state (cfg. I).

Using such a method, the number of valence electrons can be intuitively assigned to each F atom. The results are listed in Table 2. For the molecular adsorption state, there is a net charge transfer of ~ 0.2$e$ from the graphene substrate to each F atom (with reference to the number of valence electrons of free-state F, which is 7). From the transition states to the final dissociative state, the net charge transfer increases monotonically from ~ 0.3$e$ to ~ 0.6$e$, indicating the enhanced F-graphene interactions, which lead to the breaking of the F-F bond.

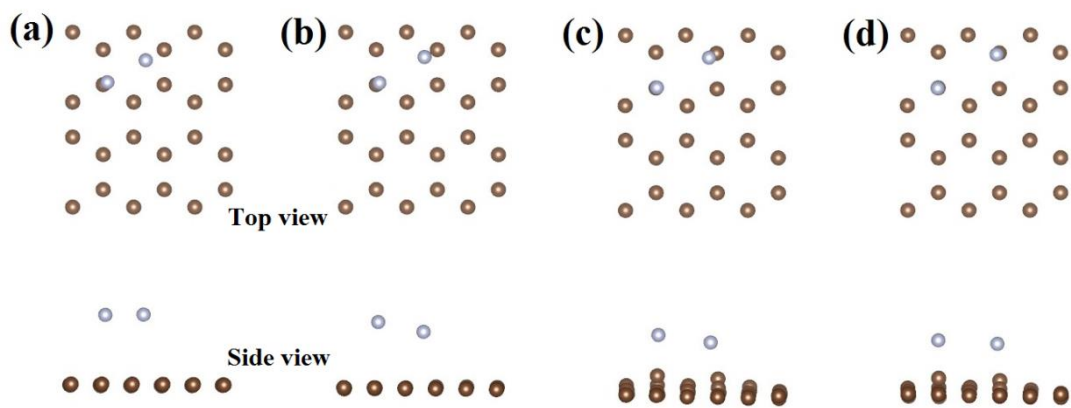

**Fig. 3.** The top (upper panels) and side views (lower panels) of the transition states A-D in Fig. 2, where a one-to-one correspondence with the panels (a)-(d) is presented. For the clarity only part of the (5×5) supercell is shown.

**Table 2.** The Bader charges assigned to the two adsorbed F atoms (labeled as F_1 and F_2) on graphene, for the adsorption Configurations V, A, B, C, D and I (Fig. 2). The corresponding distances ($Z_{F\_1}$, $Z_{F\_2}$) of the two F atoms to the graphene substrate are also listed.

|  | V | A | B | C | D | I |
|---|---|---|---|---|---|---|
| F_1 ($e$) | 7.21 | 7.28 | 7.41 | 7.52 | 7.54 | 7.59 |
| F_2 ($e$) | 7.21 | 7.29 | 7.41 | 7.54 | 7.58 | 7.58 |
| $Z_{F\_1}$ (Å) | 3.00 | 2.84 | 2.71 | 1.66 | 1.54 | 1.49 |
| $Z_{F\_2}$ (Å) | 3.18 | 2.86 | 2.23 | 1.48 | 1.49 | 1.49 |

We go further to study the diffusion of a $F_2$ molecule on the graphene surface, after its adsorption from gas phase. Our study will primarily focus on the diffusion from one most-stable configuration (Configuration I) to another most-stable one at the nearest neighboring site, which is the most probable situation in the point of view of statistical mechanics. As schematically shown in Fig. 4(a), the diffusion of a pair of F atoms (dissociative state of $F_2$) from Configuration I (labeled by the line segment $\overline{AB}$) to the neighboring most-stable configuration (labeled by the line segment $\overline{CD}$) will experience the intermediate metastable Configuration II (labeled by the line segment $\overline{BC}$). The diffusion from the configuration $\overline{AB}$ to $\overline{BC}$ (kinetic process referred to as $P$1 hereafter) involves surmounting an energy barrier of $E_{b1}$, and the diffusion from the configuration $\overline{BC}$ to $\overline{CD}$ (referred to as $P$2) needs getting across an energy barrier of $E_{b2}$. Considering the geometric symmetry of the underlying lattice of C atoms, $P$1 and $P$2 are actually reverse process of each other.

The energy landscape determined by the NEB method [20, 21] for diffusion process $P$1 is shown in Fig. 4(b), as a function of the bond length of $F_2$ molecule. The kinetic barrier $E_{b1}$ associated with process $P$1 is calculated to be ~ 1.02 eV. The energy landscape for diffusion process $P$2 is just a mirror reflection of the curve displayed in Fig. 4(b), with increasing $F_2$ bond lengths and a kinetic barrier of $E_{b2}$ ~ 0.85 eV. More generally, the diffusion between any two most-stable configurations at arbitrary separations is simply a linear combination of the diffusion pathway studied here. The values of $E_{b1}$ and $E_{b2}$ indicate that diffusion on graphene surface is nearly prohibited and the adsorption Configurations I and II are highly stable at room temperature. At higher temperatures, e.g., $T$ ~ 700 K (~ 427 ℃), the probability of surmounting $E_{b1}$ is ~ $e^{-E_{b1}/(k_B T)}$ ~ $4.53 \times 10^{-8}$, and is ~ $7.59 \times 10^{-7}$ for surmounting $E_{b2}$. Such low probability helps to understand why fluorographene can be stable up to ~ 400 ℃ [3].

To make a comparison, we have also studied the diffusion of a single F atom on the graphene surface. Our calculations show that top site adsorption is the most stable configuration for one F atom: the top site adsorption energy is ~ 0.25 eV and 0.35 eV larger than the bridge and hollow sites, respectively. Therefore, the energetically favored diffusion pathway is simply from one top site to another along the direction of C-C bonds. Shown in Fig. 5, is the calculated energy pathway for the diffusion of one F atom from the top site of a C atom (Configuration A) to its nearest neighboring top site (Configuration E). Due to the symmetry of the graphene lattice, diffusion to the other top sites at arbitrary long distance away is simply a linear combination of the pathway studied here. The kinetic barrier for the diffusion is ~ 0.25 eV, which is much smaller than the diffusion of a pair of F atoms on graphene ($E_b \geq 0.85$ eV). Such a difference implies that the interactions between the adsorbed F atoms play a key role in determining the kinetic barrier and the diffusion pathway. Therefore, except for the situation where very few F atoms are adsorbed (difficult to realize), the F-F interactions should be considered when studying their diffusion on graphene.

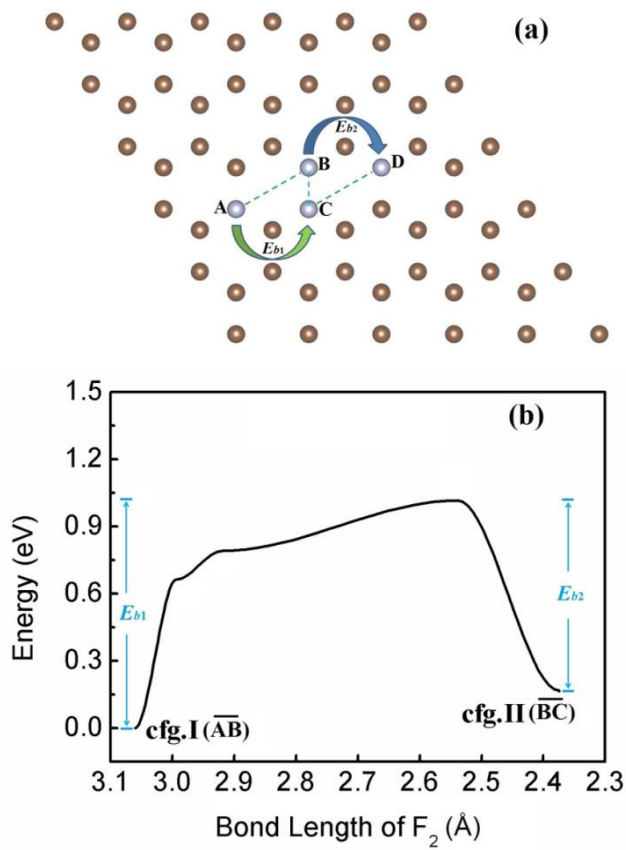

**Fig. 4.** Diffusion of $F_2$ on graphene. (a) Schematics for the transition between two neighboring most-stable configurations. (b) Calculated energy pathway for the diffusion from Configuration I to II.

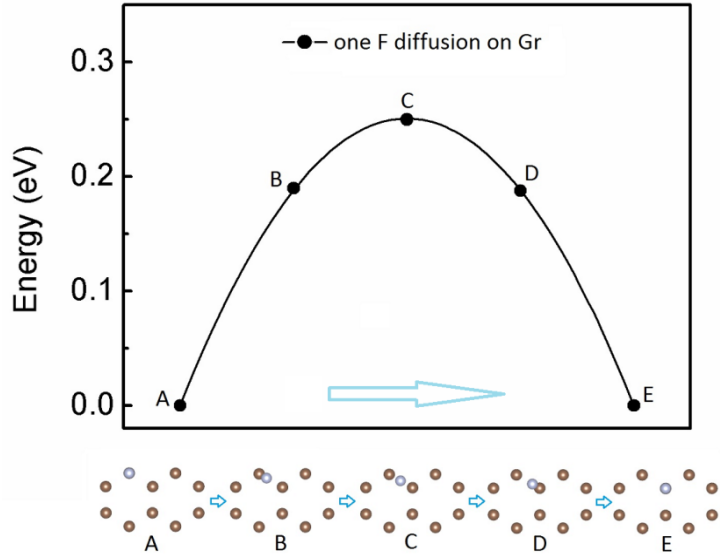

**Fig. 5.** Calculated energy pathway for the diffusion of a single F atom on graphene, from one top site (A) to another (E), via a number of transition states (B, C, D).

## 4. Conclusion

To summarize, we have studied the adsorption and diffusion of $F_2$ molecules on graphene surface by using first-principles calculations. The calculated energy pathway for the diffusion from gas phase to the most-stable surface adsorption state reveals the existence of a kinetic barrier, which stabilizes graphene in the atmosphere of $F_2$ molecules at room temperature. Meanwhile, the moderate value of the kinetic barrier (~ 0.33 eV) opens the door to activate and accelerate the reaction between graphene and $F_2$ at moderately higher temperatures. Our calculations on the energy pathway of the diffusion of $F_2$ molecules on graphene surface show the existence of high energy barriers (~ 0.8 to 1 eV), which helps to understand the stability of fluorinated graphene at room and high temperature conditions. These results shed new light to the interactions between graphene and $F_2$ molecules at the atomic level.

**Acknowledgement**

This work is financially supported by the National Natural Science Foundation of China (No. 11664003, 11474285), Natural Science Foundation of Guangxi Province (No. 2015GXNSFAA139015), and the Scientific Research and Technology Development Program of Guilin (No.2016012002). We gratefully acknowledge the crew of Center for Computational Materials Science of the Institute for Materials Research, Tohoku University for their continuous support of the SR16000 supercomputing facilities. We also thank the staff of the Supercomputing Center (Hefei Branch) of Chinese Academy of Sciences for their support of supercomputing resources.